\documentclass[twoside]{article}
\usepackage{amsmath}
\usepackage[psamsfonts]{amssymb}
\usepackage{cmmib57}
\usepackage{diagrams}
\newcommand{\bPf}{\par\vspace*{-4pt}\indent{\sc Proof.}\enskip}
\newcommand{\ePf}{\medskip}
\def\QED{\hskip0.1em\hfill\null\ \null\nobreak\hfill\kern3pt\vbox{\hrule\hbox
   {\vrule\kern1pt\vbox{\kern1.7pt\hbox{$\scriptscriptstyle{QED}$}
    \kern0.2pt}\kern1pt\vrule}\hrule}}

\def\END{\hskip0.1em\hfill\null\ \null\nobreak\hfill\kern3pt\vbox{\hrule\hbox
   {\vrule\kern1pt\vbox{\kern1.7pt\hbox{$\,\,\,\vspace{5pt}$}
    \kern0.2pt}\kern1pt\vrule}\hrule}}
\newtheorem{theorem}{Theorem}
\newtheorem{lemma}{Lemma}
\newtheorem{corollary}{Corollary}
\newtheorem{proposition}{Proposition}
\newtheorem{remark}{Remark}
\newtheorem{definition}{Definition}
\newtheorem{example}{Example}
\newcommand{\bCd}{\bEq\begin{CD}}
\newcommand{\eCd}{\end{CD}\eEq}
\newcommand{\bcd}{\beq\begin{CD}}
\newcommand{\ecd}{\end{CD}\eeq}
\newcommand{\ben}{\begin{enumerate}}
\newcommand{\een}{\end{enumerate}}
\newcommand{\bEq}{\begin{eqnarray}}
\newcommand{\eEq}{\end{eqnarray}}
\newcommand{\beq}{\begin{eqnarray*}}
\newcommand{\eeq}{\end{eqnarray*}}
\newcommand{\bDf}{\begin{definition}\em}
\newcommand{\eDf}{\end{definition}}
\newcommand{\bLm}{\begin{lemma}}
\newcommand{\eLm}{\end{lemma}}
\newcommand{\bPr}{\begin{proposition}}
\newcommand{\ePr}{\end{proposition}}
\newcommand{\bTh}{\begin{theorem}}
\newcommand{\eTh}{\end{theorem}}
\newcommand{\bCr}{\begin{corollary}}
\newcommand{\eCr}{\end{corollary}}
\newcommand{\bRm}{\begin{remark}\em}
\newcommand{\eRm}{\end{remark}}
\newcommand{\bEx}{\begin{example}\em}
\newcommand{\eEx}{\end{example}}

\newcommand{\ie}{{\em i.e$.$} }
\newcommand{\eg}{{\em e.g$.$} }
\newcommand{\R}{I\!\!R}

\newcommand{\mto}{\mapsto}

\newcommand{\der}{\partial}

\DeclareMathOperator{\im}{im}
\DeclareMathOperator{\byd}{{\raisebox{.1ex}{:}{=}}}

\newcommand{\ucar}[1]{\underset{#1}{\times}}

\newcommand{\owed}[1]{\overset{#1}{\wedge}}

\newcommand{\balp}{\boldsymbol{\alp}}

\newcommand{\cA}{\mathcal{A}}

\newcommand{\cC}{\mathcal{C}}

\newcommand{\cE}{\mathcal{E}}

\newcommand{\cH}{\mathcal{H}}

\newcommand{\cJ}{\mathcal{J}}

\newcommand{\cL}{\mathcal{L}}

\newcommand{\cT}{\mathcal{T}}

\newcommand{\by}{\boldsymbol{y}}

\newcommand{\bF}{\boldsymbol{F}}
\newcommand{\bG}{\boldsymbol{G}}

\newcommand{\bP}{\boldsymbol{P}}

\newcommand{\bW}{\boldsymbol{W}}
\newcommand{\bX}{\boldsymbol{X}}
\newcommand{\bY}{\boldsymbol{Y}}

\newcommand{\sub}{\subset}

\newcommand{\wed}{\wedge}
\newcommand{\com}{\!\circ\!}
\newcommand{\ten}{\!\otimes\!}

\newcommand{\alp}{\alpha}
\newcommand{\bet}{\beta}
\newcommand{\gam}{\gamma}
\newcommand{\del}{\delta}

\newcommand{\zet}{\zeta}

\newcommand{\lam}{\lambda}
\newcommand{\sig}{\sigma}

\newcommand{\ome}{\omega}

\newcommand{\Lam}{\Lambda}

\newcommand{\Ome}{\Omega}

\newcommand{\vartht}{\vartheta}

\newcommand{\For}{{\Lambda}}
\newcommand{\Con}{{\mathcal{C}}}
\newcommand{\Hor}{{\mathcal{H}}}
\newcommand{\Var}{{\mathcal{V}}}
\newcommand{\Thd}{{\Theta}}

\title{\large{{\bf Generalized Bianchi identities in gauge-natural field theories and the 
  curvature of variational principles}\thanks{Partially supported by
GNFM of INdAM, MIUR (PRIN 2003) and University of Torino.}}}
\author{{\normalsize M. Francaviglia, M.
Palese and E. Winterroth}
\\{\footnotesize Department of Mathematics,
University of Torino}
\\{\footnotesize via C. Alberto 10, 10123 Torino, Italy}\\ 
{\footnotesize e--mails: 
{\sc francaviglia@dm.unito.it, palese@dm.unito.it, ekkehart@dm.unito.it}}}
\date{}
\begin{document}

\maketitle

\begin{abstract}

  By resorting to Noether's Second Theorem, we relate the generalized
  Bianchi identities for Lagrangian field theories on gauge-natural bundles 
  with the kernel of the associated gauge-natural Jacobi morphism. A suitable 
  definition of the curvature of gauge-natural variational principles can be 
  consequently formulated in terms of the Hamiltonian connection canonically 
associated with a generalized Lagrangian obtained by contracting
field equations. 
                                         
\medskip


\end{abstract}

\section{Introduction}

To investigate conservation laws for covariant field theories, see \eg 
\cite{Ber58} and references quoted therein, 
the general problem has been tackled of coherently
defining the lifting of infinitesimal transformations of the basis manifolds to the bundle
of fields (namely bundles of tensor fields or tensor densities which could be obtained as
suitable representations of the action of infinitesimal space-time transformations on frame
bundles of a given order \cite{PaTe77}). Such theories were also called geometric or
{\em natural} \cite{Tra67}. An important generalization of natural theories to gauge fields theories passed
through 
the concept of jet prolongation of a principal bundle and the
introduction of a very important geometric construction, namely the {\em gauge-natural} bundle
functor \cite{Ec81,KMS93}. 

Within the above mentioned general program generalized Bianchi identities for geometric field
theories were introduced to get (after an  integration by parts procedure) a consistent
equation involving divergences within the first variation formula. It was
also stressed that in the general theory of relativity these identities
coincide with the  contracted Bianchi identities for the curvature tensor of
the pseudo-Riemannian metric. 

Our general framework is the calculus of variations on finite order fibered 
bundles. Fibered bundles will be assumed to be 
{\em gauge-natural bundles} (\ie jet prolongations of fiber bundles associated 
to some gauge-natural prolongation of a principal bundle $\bP$ 
\cite{Ec81,KMS93}). 
Such geometric structures have been widely recognized to suitably describe 
so-called gauge-natural field theories, \ie physical theories in which
right-invariant infinitesimal automorphisms of the structure bundle $\bP$  
uniquely define the transformation laws of the fields themselves (see {\em e.g.} 
\cite{FaFr03} and references quoted therein).
We 
shall in particular consider {\em finite order variational sequences
on gauge-natural bundles}.
 The gauge-natural lift enables one  to define the generalized gauge-natural Jacobi morphism
where the {\em variation vector fields} 
are Lie derivatives 
of sections of the gauge-natural bundle with respect to gauge-natural lifts of infinitesimal 
automorphisms of the underlying principal bundle $\bP$. 
Within such a picture, as a consequence of the Second Noether Theorem, it is possible to  relate the generalized Bianchi
morphism to the second variation of the Lagrangian and then to the associated Jacobi morphism
\cite{FrPa00,FrPa01,PaWi03,PaWi04}. 
In the 
case of geodesics in a Riemannian manifold vector fields which 
make the second variation to vanish identically modulo boundary terms are 
called {\em Jacobi fields} and they are solutions of a second--order 
differential equation known as {\em Jacobi equation} (of geodesics).
The notion of Jacobi equation as an outcome of the second variation is in 
fact fairly more general: formulae for the second 
variation of a Lagrangian functional and generalized Jacobi equations along critical sections have been 
already considered (see, \eg the results of \cite{Vari} and classical refrences quoted therein).
 In this paper we show that, as a consequence of the gauge-natural invariance of the
Lagrangian, there exists a covariantly conserved current associated with the contraction of the Euler--Lagrange morphism with a gauge-natural
Jacobi vector field. Such a conserved current can be considered as a Hamiltonian for the Lagrangian corresponding to such a contraction. The
curvature of the variational principle can be then defined as the curvature of the corresponding Hamiltonian connection.

\section{Finite order jets of gauge-natural bundles and variational sequences}

In this Section we recall some basic facts about jet spaces.
We introduce jet spaces of a
fibered manifold and the sheaves of forms on the $s$--th order
jet space. Moreover, we recall the notion of horizontal and vertical
differential \cite{KMS93,Sau89}.

Our framework is a fibered manifold $\pi : \bY \to \bX$,
with $\dim \bX = n$ and $\dim \bY = n+m$.

For $s \geq q \geq 0$ integers we are concerned with the $s$--jet space $J_s\bY$ of $s$--jet prolongations of (local) sections
of $\pi$; in particular, we set $J_0\bY \equiv \bY$. We recall the natural fiberings
$\pi^s_q: J_s\bY \to J_q\bY$, $s \geq q$, $\pi^s: J_s\bY \to \bX$, and,
among these, the {\em affine\/} fiberings $\pi^{s}_{s-1}$.
We denote by $V\bY$ the vector subbundle of the tangent
bundle $T\bY$ of vectors on $\bY$ which are vertical with respect
to the fibering $\pi$.

Greek
indices $\sig ,\mu ,\dots$ run from $1$ to $n$ and they label basis
coordinates, while
Latin indices $i,j,\dots$ run from $1$ to $m$ and label fibre coordinates,
unless otherwise specified. 
We denote multi--indices of dimension $n$ by boldface Greek letters such as
$\balp = (\alp_1, \dots, \alp_n)$, with $0 \leq \alp_\mu$,
$\mu=1,\ldots,n$; by an abuse
of notation, we denote by $\sig$ the multi--index such that
$\alp_{\mu}=0$, if $\mu\neq \sig$, $\alp_{\mu}= 1$, if
$\mu=\sig$.
We also set $|\balp| \byd \alp_{1} + \dots + \alp_{n}$ and $\balp ! \byd
\alp_{1}! \dots \alp_{n}!$.
The charts induced on $J_s\bY$ are denoted by $(x^\sig,y^i_{\balp})$, with $0
\leq |\balp| \leq s$; in particular, we set $y^i_{\bf{0}}
\equiv y^i$. The local vector fields and forms of $J_s\bY$ induced by
the above coordinates are denoted by $(\der^{\balp}_i)$ and $(d^i_{\balp})$,
respectively.

For $s\geq 1$, we consider the natural complementary fibered
morphisms over $J_s\bY \to J_{s-1}\bY$ (see \cite{Kru90,Kru93,Vit98}):
\beq
\mathcal{D} : J_s\bY \ucar{\bX} T\bX \to TJ_{s-1}\bY \,,
\qquad
\vartht : J_{s}\bY \ucar{J_{s-1}\bY} TJ_{s-1}\bY \to VJ_{s-1}\bY \,,
\eeq
with coordinate expressions, for $0 \leq |\balp| \leq s-1$, given by
\beq
\mathcal{D} &= d^\lam\ten {\mathcal{D}}_\lam = d^\lam\ten
(\der_\lam + y^j_{\balp+\lam}\der_j^{\balp}) \,,
\vartht &= \vartht^j_{\balp}\ten\der_j^{\balp} =
(d^j_{\balp}-y^j_{{\balp}+\lam}d^\lam)
\ten\der_j^{\balp} \,.
\eeq

The morphisms above induce the following natural splitting (and its dual):
\bEq
\label{jet connection}
J_{s}\bY\ucar{J_{s-1}\bY}T^*J_{s-1}\bY =\left(
J_s\bY\ucar{J_{s-1}\bY}T^*\bX\right) \oplus\cC^{*}_{s-1}[\bY]\,,
\eEq
where $\cC^{*}_{s-1}[\bY] \byd \im \vartht_s^*$ and
$\vartht_s^* : J_s\bY \ucar{J_{s-1}\bY} V^*J_{s-1}\bY \to
J_s\bY \ucar{J_{s-1}\bY} T^*J_{s-1}\bY \,$.

If $f: J_{s}\bY \to \R$ is a function, then we set
$D_{\sig}f$ $\byd \mathcal{D}_{\sig} f$,
$D_{\balp+\sig}f$ $\byd D_{\sig} D_{\balp}f$, where $D_{\sig}$ is
the standard {\em formal derivative}.
Given a vector field $\Xi : J_{s}\bY \to TJ_{s}\bY$, the splitting
\eqref{jet connection} yields $\Xi \, \com \, \pi^{s+1}_{s} = \Xi_{H} + \Xi_{V}$
where, if $\Xi = \Xi^{\gam}\der_{\gam} + \Xi^i_{\balp}\der^{\balp}_i$, then we
have $\Xi_{H} = \Xi^{\gam}D_{\gam}$ and
$\Xi_{V} = (\Xi^i_{\balp} - y^i_{\balp + \gam}\Xi^{\gam}) 
\der^{\balp}_{i}$. We shall call $\Xi_{H}$ and $\Xi_{V}$ the 
horizontal and the vertical part of $\Xi$, respectively.

The splitting
\eqref{jet connection} induces also a decomposition of the
exterior differential on $\bY$,
$(\pi^{s}_{s-1})^*\com \,d = d_H + d_V$, where $d_H$ and $d_V$
are defined to be the {\em horizontal\/} and {\em vertical differential\/}.
The action of $d_H$ and $d_V$ on functions and $1$--forms
on $J_s\bY$ uniquely characterizes $d_H$ and $d_V$ (see, {\em e.g.},
\cite{Sau89,Vit98} for more details).
A {\em projectable vector field\/} on $\bY$ is defined to be a pair
$(u,\xi)$, where $u:\bY \to T\bY$ and $\xi: \bX \to T\bX$
are vector fields and $u$ is a fibered morphism over $\xi$.
If there is no danger of confusion, we will denote simply by $u$ a
projectable vector field $(u,\xi)$.
A projectable vector field $(u,\xi)$
can be conveniently prolonged to a projectable vector field
$(j_{s}u, \xi)$; coordinate expression can be found \eg in 
\cite{Kru90,Sau89,Vit98}.

\subsection{Gauge-natural bundles}

Let $\bP\to\bX$ be a principal bundle with structure group $\bG$.
Let $r\leq k$ be integers and $\bW^{(r,k)}\bP$ $\byd$ $J_{r}\bP\ucar{\bX}L_{k}(\bX)$, 
where $L_{k}(\bX)$ is the bundle of $k$--frames 
in $\bX$ \cite{Ec81,KMS93}, $\bW^{(r,k)}\bG \byd J_{r}\bG\odot GL_{k}(n)$
the semidirect product with respect to the action of $GL_{k}(n)$ 
on $J_{r}\bG$ given by the 
jet composition and $GL_{k}(n)$ is the group of $k$--frames 
in $\R^{n}$. Here we denote by $J_{r}\bG$ the space of $(r,n)$-velocities on $\bG$.
The bundle $\bW^{(r,k)}\bP$ is a principal bundle over $\bX$ with structure group
$\bW^{(r,k)}\bG$.
Let $\bF$ be any manifold and $\zet:\bW^{(r,k)}\bG\ucar{}\bF\to\bF$ be 
a left action of $\bW^{(r,k)}\bG$ on $\bF$. There is a naturally defined 
right action of $\bW^{(r,k)}\bG$ on $\bW^{(r,k)}\bP \times \bF$ so that
 we can associate in a standard way
to $\bW^{(r,k)}\bP$ the bundle, on the given basis $\bX$,
$\bY_{\zet} \byd \bW^{(r,k)}\bP\times_{\zet}\bF$.

\bDf
We say $(\bY_{\zet},\bX,\pi_{\zet};\bF,\bG)$ to be the 
{\em gauge-natural bundle} of order 
$(r,k)$ associated to the principal bundle $\bW^{(r,k)}\bP$ 
by means of the left action $\zet$ of the group 
$\bW^{(r,k)}\bG$ on the manifold $\bF$ \cite{Ec81,KMS93}. 
\END\eDf

\bRm
A principal automorphism $\Phi$ of $\bW^{(r,k)}\bP$ induces an 
automorphism of the gauge-natural bundle by:
\bEq
\Phi_{\zet}:\bY_{\zet}\to\bY_{\zet}: [(j^{x}_{r}\gam,j^{0}_{k}t), 
\hat{f}]_{\zet}\mto [\Phi(j^{x}_{r}\gam,j^{0}_{k}t), 
\hat{f}]_{\zet}\,, 
\eEq
where $\hat{f}\in \bF$ and $[\cdot, \cdot]_{\zet}$ is the equivalence class
induced by the action $\zet$.
\END\eRm
\bDf
We define the {\em vector}
bundle over $\bX$ of right--invariant infinitesimal automorphisms of $\bP$
by setting $\cA = T\bP/\bG$. 

We also define the {\em vector} bundle  over $\bX$ of right invariant 
infinitesimal automorphisms of $\bW^{(r,k)}\bP$ by setting 
$\cA^{(r,k)} \byd T\bW^{(r,k)}\bP/\bW^{(r,k)}\bG$ ($r\leq k$).
\END\eDf

Denote by $\cT_{\bX}$ and $\cA^{(r,k)}$ the sheaf of
vector fields on $\bX$ and the sheaf of right invariant vector fields 
on $\bW^{(r,k)}\bP$, respectively. A functorial mapping $\mathfrak{G}$ is defined 
which lifts any right--invariant local automorphism $(\Phi,\phi)$ of the 
principal bundle $W^{(r,k)}\bP$ into a unique local automorphism 
$(\Phi_{\zet},\phi)$ of the associated bundle $\bY_{\zet}$. 
Its infinitesimal version associates to any $\bar{\Xi} \in \cA^{(r,k)}$,
projectable over $\xi \in \cT_{\bX}$, a unique {\em projectable} vector field 
$\hat{\Xi} \byd \mathfrak{G}(\bar{\Xi})$ (the gauge-natural lift) on $\bY_{\zet}$ in the 
following way:
\bEq
\mathfrak{G} : \bY_{\zet} \ucar{\bX} \cA^{(r,k)} \to T\bY_{\zet} \,:
(\by,\bar{\Xi}) \mto \hat{\Xi} (\by) \,,
\eEq
where, for any $\by \in \bY_{\zet}$, one sets: $\hat{\Xi}(\by)=
\frac{d}{dt} [(\Phi_{\zet \,t})(\by)]_{t=0}$,
and $\Phi_{\zet \,t}$ denotes the (local) flow corresponding to the 
gauge-natural lift of $\Phi_{t}$.

This mapping fulfils the following properties (see \cite{KMS93}):
\begin{enumerate}
\item $\mathfrak{G}$ is linear over $id_{\bY_{\zet}}$;
\item we have $T\pi_{\zet}\circ\mathfrak{G} = id_{T\bX}\circ 
\bar{\pi}^{(r,k)}$, 
where $\bar{\pi}^{(r,k)}$ is the natural projection
$\bY_{\zet}\ucar{\bX} 
\cA^{(r,k)} \to T\bX$;
\item for any pair $(\bar{\Lam},\bar{\Xi})$ $\in$
$\cA^{(r,k)}$, we have
$\mathfrak{G}([\bar{\Lam},\bar{\Xi}]) = [\mathfrak{G}(\bar{\Lam}), \mathfrak{G}(\bar{\Xi})]$.
\end{enumerate}

\subsection{Lie derivative}

\bDf
Let $\gam$ be a (local) section of $\bY_{\zet}$, $\bar{\Xi}$ 
$\in \cA^{(r,k)}$ and $\hat\Xi$ its gauge-natural lift. 
Following \cite{KMS93} we
define the {\em 
generalized Lie derivative} of $\gam$ along the vector field 
$\hat{\Xi}$ to be the (local) section $\pounds_{\bar{\Xi}} \gam : \bX \to V\bY_{\zet}$, 
given by
$\pounds_{\bar{\Xi}} \gam = T\gam \circ \xi - \hat{\Xi} \circ \gam$.\END
\eDf

\bRm\label{lie}
The Lie derivative operator acting on sections of gauge-natural 
bundles satisfies the following 
properties:
\begin{enumerate}\label{lie properties}
\item for any vector field $\bar{\Xi} \in \cA^{(r,k)}$, the 
mapping $\gam \mto \pounds_{\bar{\Xi}}\gam$ 
is a first--order quasilinear differential operator;
\item for any local section $\gam$ of $\bY_{\zet}$, the mapping 
$\bar{\Xi} \mto \pounds_{\bar{\Xi}}\gam$ 
is a linear differential operator;
\item we can regard $\pounds_{\bar{\Xi}}: J_{1}\bY_{\zet} \to V\bY_{\zet}$ 
as a morphism over the
basis $\bX$. By using the canonical 
isomorphisms $VJ_{s}\bY_{\zet}\simeq J_{s}V\bY_{\zet}$ for all $s$, we have
$\pounds_{\bar{\Xi}}[j_{s}\gam] = j_{s} [\pounds_{\bar{\Xi}} \gam]$,
for any (local) section $\gam$ of $\bY_{\zet}$ and for any (local) 
vector field $\bar{\Xi}\in \cA^{(r,k)}$. 

\end{enumerate}
\END\eRm

\subsection{Variational sequences} 

For the sake of simplifying notation, sometimes, we will omit the subscript $\zet$, so 
that all our considerations shall refer to $\bY$ as a gauge-natural 
bundle as defined above.

\bRm
According to \cite{Kru90,Kru93,Vit98}, the fibered splitting
\eqref{jet connection} yields the {\em sheaf splitting}
$\Hor^{p}_{(s+1,s)}$ $=$ $\bigoplus_{t=0}^p$
$\Con^{p-t}_{(s+1,s)}$ $\wed\Hor^{t}_{s+1}$, which restricts to the inclusion
$\For^{p}_s$ $\sub$ $\bigoplus_{t=0}^{p}$
$\Con^{p-t}{_s}\wed\Hor^{t,}{_{s+1}^{h}}$,
where $\For, \Hor, \Con$ are respectively sheaves of differential, horizontal and contact forms in the sense of \cite{Kru90},
$\Hor^{p,}{_{s+1}^{h}}$
$\byd$
$h(\For^{p}_s)$ for
$0 < p\leq n$ and the surjective map
$h$ is defined to be the restriction to $\For^{p}_{s}$ of the projection of
the above splitting onto the non--trivial summand with the highest
value of $t$ (see \eg \cite{PaWi03,PaWi04} for notation).\END
\eRm

By an abuse of notation, let us denote by $d\ker h$ the sheaf
generated by the presheaf $d\ker h$ in the standard way.
We set $\Thd^{*}_{s}$ $\byd$ $\ker h$ $+$
$d\ker h$.

In \cite{Kru90} 
it was proved that the following sequence is an exact resolution of the constant sheaf $\R_{\bY}$ over $\bY$:
\beq
\diagramstyle[size=1.3em]
\begin{diagram}
0 & \rTo & \R_{Y} & \rTo & \For^{0}_s & \rTo^{\cE_{0}} &
\For^{1}_s/\Thd^{1}_s & \rTo^{\cE_{1}} & \For^{2}_s/\Thd^{2}_s & \rTo^{\cE_{2}} &
\dots & \rTo^{\cE_{I-1}} & \For^{I}_s/\Thd^{I}_s & \rTo^{\cE_{I}} &
\For^{I+1}_s & \rTo^{d} & 0
\end{diagram}
\eeq

\bDf
The above sequence, where the highest integer $I$ depends on the dimension
of the fibers of $J_{s}\bY \to \bX$ (see, in particular, \cite{Kru90}), is said to be the $s$--th order
{\em variational sequence\/} associated with the fibered manifold
$\bY\to\bX$.
\END
\eDf

For practical purposes we 
shall limit ourself to consider the truncated variational sequence: 
\beq
\diagramstyle[size=1.3em]
\begin{diagram}
0 &\rTo & \R_{Y} &\rTo & \Var^{0}_s & \rTo^{\cE_0} &
\Var^{1}_{s} & \rTo^{\cE_{1}} & \dots  & \rTo^{\cE_{n}} &
\Var^{n+1}_{s}  & \rTo^{\cE_{n+1}} & \cE_{n+1}(\Var^{n+1}_{s})  
& \rTo^{\cE_{n+2}} & 
0 \,,
\end{diagram}
\eeq
where, following \cite{Vit98}, the sheaves $\Var^{p}_{s}\byd 
\Con^{p-n}_{s}\wed\Hor^{n,}{_{s+1}^h}/h(d\ker h)$ with $0\leq p\leq n+2$ are 
suitable representations of the corresponding quotient 
sheaves in the variational sequence by means of sheaves of sections of tensor
bundles.

Let $\alp\in\Con^{1}_s\wed\Hor^{n,}{_{s+1}^h} 
\sub \Var^{n+1}_{s+1}$. Then there is a unique pair of
sheaf morphisms (\cite{Kol83,KoVi03,Vit98})
\bEq\label{first variation}
E_{\alp} \in \Con^{1}_{(2s,0)}\wed\Hor^{n,}{_{2s+1}^{h}} \,,
\qquad
F_{\alp} \in \Con^{1}_{(2s,s)} \wed \Hor^{n,}{_{2s+1}^h} \,,
\eEq
such that 
$(\pi^{2s+1}_{s+1})^*\alp=E_{\alp}-F_{\alp}$
and $F_\alp$ is {\em locally} of the form $F_{\alp} = d_{H}p_{\alp}$, with $p_{\alp}
\in \Con^{1}_{(2s-1,s-1)}\wed\Hor^{n-1}{_{2s}}$.

Let $\eta\in\Con^{1}_{s}\wed\Con^{1}_{(s,0)}\wed\Hor^{n,}{_{s+1}^{h}}\sub 
\Var^{n+2}_{s+1}$;
then there is a unique morphism 
$$
K_{\eta} \in \Con^{1}_{(2s,s)}\otimes\Con^{1}_{(2s,0)}\wed\Hor^{n,}{_{2s+1}^{h}}
$$
such that, for all $\Xi:\bY\to V\bY$,
$
E_{{j_{s}\Xi}\rfloor \eta} = C^{1}_{1} (j_{2s}\Xi\ten K_{\eta})$,
where $C^1_1$ stands for tensor
contraction on the first factor and $\rfloor$ denotes inner product (see \cite{KoVi03,Vit98}). 
Furthermore, there is a unique pair of sheaf morphisms
\bEq\label{second}
H_{\eta} \in 
\Con^{1}_{(2s,s)}\wed\Con^{1}_{(2s,0)}\wed\Hor^{n,}{_{2s+1}^{h}} \,,
\quad
G_{\eta} \in \Con^{2}_{(2s,s)}\wed\Hor^{n,}{_{2s+1}^{h}} \,,
\eEq
such that 
${(\pi^{2s+1}_{s+1})}^*\eta = H_{\eta} - G_{\eta}$ and $H_{\eta} 
= \frac{1}{2} \, A(K_{\eta})$,
where $A$ stands for antisymmetrisation.
Moreover, $G_{\eta}$ is {\em locally} of the type $G_{\eta} = d_H q_{\eta}$, 
where 
$q_{\eta} \in \Con^{2}_{(2s-1,s-1)}\wed\Hor^{n-1}{_{2s}}$; hence 
$[\eta]=[H_{\eta}]$ \cite{KoVi03,Vit98}. 

\bRm
A section $\lam\in\Var^{n}_s$ is just a Lagrangian of order 
$(s+1)$ of 
the standard literature. 
Furthermore
$\cE_{n}(\lam) \in \Var^{n+1}_{s}$ coincides with the standard higher 
order Euler--Lagrange morphism associated with $\lam$.
\END
\eRm

\section{Variations and generalized Jacobi morphisms}

We recall some previous results concerning 
the representation of {\em generalized gauge-natural Jacobi morphisms} in
variational sequences and their relation with the second variation of a 
generalized gauge-natural invariant Lagrangian.
We consider {\em formal variations} of a morphism as  
{\em multiparameter deformations} and relate
 the second variation of 
the Lagrangian $\lam$ to the
Lie derivative of the associated Euler--Lagrange morphism and to the generalized
Bianchi morphism; see \cite{PaWi03} for details.
\medskip

\bDf\label{var}
Let $\pi: \bY\to \bX$ be any bundle and let  $\alp: J_{s}\bY\to
\owed{p}T^*J_{s}\bY$ and $L_{j_{s}\Xi_{k}}$ be the Lie derivative 
operator acting on differential fibered morphism.
Let 
 $\Xi_{k}$, $1\leq k\leq i$, be (vertical) variation 
vector fields on $\bY$ in the sense of \cite{FrPa00,FrPa01,FPV02,PaWi03}. We define the
$i$--th formal variation of the morphism $\alp$ to be the operator:
$\del^{i} \alp = L_{j_{s}\Xi_{1}} \ldots L_{j_{s}\Xi_{i}} \alp$.
\eDf

\bRm
Let $\alp\in (\Var^{n}_{s})_{\bY}$. 

\noindent We have $
\del^{i}[\alp]\byd [\del^{i}\alp]$  $=$ $[L_{\Xi_{i}} \ldots
L_{\Xi_{1}}\alp]$ $=$ $\cL_{\Xi_{i}} \ldots 
\cL_{\Xi_{1}}[\alp]$.\END
\eRm

\bDf
We call the operator $\del^{i}$ the {\em $i$--th vertical variational
derivative}. \END
\eDf

As a consequence of the Second Noether Theorem (see \eg \cite{FPV98a,PaWi03}), the following
characterization of the second variational vertical derivative of a generalized Lagrangian 
in the variational sequence holds true.

\bPr\label{x}
Let $\lam\in (\Var^{n}_{s})_{\bY}$ and let $\Xi$ be a variation vector 
field; then we have
\bEq
\del^{2}\lam = [\cE_{n}(j_{2s}\Xi \rfloor h\del\lam)
+C^{1}_{1} (j_{2s}\Xi \ten K_{hd\del\lam})] \,.
\eEq
\ePr

\subsection{Generalized {\em gauge-natural} Jacobi morphisms}

Let now specify $\bY$ to be a gauge-natural bundle and let $\hat{\Xi}\equiv \mathfrak{G}(\bar{\Xi})$ be a variation vector field associated to
some
$\bar{\Xi}\in \cA^{r,k}$. Let us consider
$j_{s}{\hat{\Xi}}_{V}$,
\ie the vertical part according to the splitting \eqref{jet connection}. We shall denote by $j_{s}\bar{\Xi}_{V}$ the induced section of the
vector bundle
$\cA^{(r+s,k+s)}$. The set of all sections of this kind defines a vector subbundle of $J_{s}\cA^{(r,k)}$, which by a slight abuse of notation
(since we are speaking  about vertical parts with respect to the splitting \eqref{jet 
connection}), we
shall denote by 
$VJ_{s}\cA^{(r,k)}$. 

By applying an abstract result due to Kol\'a\v r, see
\cite{Kol83}, concerning a global decomposition formula for vertical morphisms, and by using Proposition \ref{x} we can prove the following.

\bLm
Let $\lam$ be a Lagrangian and $\hat{\Xi}_V$ a variation vecrtor 
field. Let us set $\chi(\lam,\mathfrak{G}(\bar{\Xi})_{V})\byd
C^{1}_{1} (j_{2s}\hat{\Xi}\ten 
K_{hd\cL_{j_{2s}\bar{\Xi}_V}\lam})\equiv E_{j_{s}\hat{\Xi}\rfloor
hd\cL_{j_{2s+1}\bar{\Xi}_V}\lam}$. Let
$D_{H}$ be the horizontal differential on 
$J_{4s}\bY_{\zet}\ucar{\bX}VJ_{4s}\cA^{(r,k)}$.
Then we have:
\bEq\label{chi}
(\pi^{4s+1}_{2s+1})^{*}\chi(\lam,\mathfrak{G}(\bar{\Xi})_{V}) = 
E_{\chi(\lam,\mathfrak{G}(\bar{\Xi})_{V})} +
F_{\chi(\lam,\mathfrak{G}(\bar{\Xi})_{V})}\,,
\eEq
where
\beq
E_{\chi(\lam,\mathfrak{G}(\bar{\Xi})_{V}}: 
J_{4s}\bY_{\zet}\ucar{\bX}VJ_{4s}\cA^{(r,k)} \to
\Con^{*}_{0}[\cA^{(r,k)}]\ten\Con^{*}_{0}[\cA^{(r,k)}]\wed 
(\owed{n}T^{*}\bX) \,,
\eeq
and locally $F_{\chi(\lam,\mathfrak{G}(\bar{\Xi})_{V})} =
D_{H}M_{\chi(\lam,\mathfrak{G}(\bar{\Xi})_{V})}$ with
\beq
M_{\chi(\lam,\mathfrak{G}(\bar{\Xi})_{V})}:J_{4s}\bY_{\zet}\ucar{\bX}VJ_{4s}\cA^{(r,k)}\to 
\Con^{*}_{2s-1}[\cA^{(r,k)}]\ten \Con^{*}_{0}[\cA^{(r,k)}]\wed
(\owed{n-1}T^{*}\bX)\,.
\eeq
\eLm

\bPf
As a consequence of
linearity properties
of both $\chi(\lam, \Xi)$ and
  the Lie derivative operator $\pounds$ we have 
$\chi(\lam,\mathfrak{G}(\bar{\Xi})_{V}):
J_{2s}\bY_{\zet}\ucar{\bX}VJ_{2s}\cA^{(r,k)} \to
\Con^{*}_{2s}[\cA^{(r,k)}]\ten
\Con^{*}_{0}[\cA^{(r,k)}] \wed (\owed{n}T^{*}\bX)$ and
$D_{H}\chi(\lam,\mathfrak{G}(\bar{\Xi})_{V})$ $=$ $0$.
Thus Kol\'a\v r's decomposition formula can be applied. 
\QED
\ePf

\bDf
We call the morphism $\cJ(\lam,\mathfrak{G}(\bar{\Xi})_{V})$ $\byd$
$E_{\chi(\lam,\mathfrak{G}(\bar{\Xi})_{V})}$  the {\em gauge-natural generalized Jacobi 
morphism} associated with the Lagrangian $\lam$ and the variation vector field $\bar{\Xi}$ 
(it depends on the gauge-natural lift
$\mathfrak{G}(\bar{\Xi})_{V}$). Coordinate expressions can be found \eg in \cite{PaWi03}.
\END\eDf

The morphism $\cJ(\lam,\mathfrak{G}(\bar{\Xi})_{V})$ is a 
{\em linear} morphism with respect to the projection 
$J_{4s}\bY_{\zet}\ucar{\bX}VJ_{4s}\cA^{(r,k)}\to J_{4s}\bY_{\zet}$.
We have then (see \cite{PaWi03}) the following.

\bTh\label{comparison}
Let $\del^{2}_{\mathfrak{G}}\lam$ be the variation of $\lam$ with respect to vertical parts
of gauge-natural lifts of infinitesimal principal automorphisms. We have:
\bEq
\mathfrak{G}(\bar{\Xi})_{V}\rfloor
\cE_{n}(\mathfrak{G}(\bar{\Xi})_{V}\rfloor\cE_{n}(\lam))
=
\del^{2}_{\mathfrak{G}}\lam 
=
\cE_{n}(\mathfrak{G}(\bar{\Xi})_{V}\rfloor 
h(d\del\lam))\,. 
\eEq
\eTh

\bPr
\label{symmetry of L}
Let $\lam \in \Var^{n}_{s}$ be a gauge-natural Lagrangian and 
$(\hat{\Xi},\xi)$ 
a gauge-natural symmetry of $\lam$. Then we have
$0 = - \pounds_{\bar{\Xi}} \rfloor \cE_{n}(\lam) 
+d_{H}(-j_{s}\pounds_{\bar{\Xi}} 
\rfloor p_{d_{V}\lam}+ \xi \rfloor \lam)$.
\ePr (See \eg \cite{Tra67} for geometric field theories and, in particular, 
\cite{FPV98a,PaWi04}).

\subsection{Bianchi morphism and curvature}

Generalized Bianchi identities for field theories
are necessary and (locally) sufficient conditions for the Noether conserved current
$\epsilon=-j_{s}\pounds_{\bar{\Xi}} 
\rfloor p_{d_{V}\lam}+ \xi \rfloor \lam$ to be not only closed but also the divergence of a skew-symmetric (tensor) density
along solutions of the  Euler--Lagrange equations (see \eg \cite{Tra67}).

Let now $\lam$ be a gauge-natural Lagrangian. We set
\bEq
\ome (\lam,\mathfrak{G}(\bar{\Xi})_{V}) \equiv  \pounds_{\bar{\Xi}} \rfloor \cE_{n} (\lam): J_{2s}\bY_{\zet}
\to \Con_{2s}^{*}[\cA^{(r,k)}]\ten \Con_{0}^{*}[\cA^{(r,k)}]\wed (\owed{n} 
T^{*}\bX) \,.
\eEq
The morphism $\ome (\lam,\mathfrak{G}(\bar{\Xi})_{V})$ so defined is a generalized Lagrangian associated with the field equations of the
original Lagrangian $\lam$ . It has been considered in applications \eg in General Relativity (see \cite{FFR03} and references quoted therein).
We have $D_{H}\ome(\lam,\mathfrak{G}(\bar{\Xi})_{V})= 0$ and by the linearity of
$\pounds$ we can regard $\ome(\lam,\mathfrak{G}(\bar{\Xi})_{V})$ as the extended morphism 
$\ome(\lam,\mathfrak{G}(\bar{\Xi})_{V}): J_{2s}\bY_{\zet} \ucar{\bX} VJ_{2s}\cA^{(r,k)})
\to \Con_{2s}^{*}[\cA^{(r,k)}]\ten\Con_{2s}^{*}[\cA^{(r,k)}]\ten\Con_{0}^{*}[\cA^{(r,k)}]\wed
(\owed{n}T^{*}\bX)$. Thus we can state the following \cite{PaWi03}.

\bLm\label{kol}
Let $\ome(\lam,\mathfrak{G}(\bar{\Xi})_{V})$ be as above. Then we have {\em globally}
\beq
(\pi^{4s+1}_{s+1})^{*}\ome(\lam,\mathfrak{G}(\bar{\Xi})_{V}) = \bet(\lam,\mathfrak{G}(\bar{\Xi})_{V}) +
F_{\ome(\lam,\mathfrak{G}(\bar{\Xi})_{V})}\,,
\eeq 
where
\bEq\label{bianchi}
&\bet(\lam,\mathfrak{G}(\bar{\Xi})_{V})\equiv
E_{\ome(\lam,\mathfrak{G}(\bar{\Xi})_{V})} : \\
& J_{4s}\bY_{\zet} \ucar{\bX} VJ_{4s}\cA^{(r,k)} \to 
\Con_{2s}^{*}[\cA^{(r,k)}]\ten\Con_{0}^{*}[\cA^{(r,k)}]\ten\Con_{0}^{*}[\cA^{(r,k)}]\wed
(\owed{n}T^{*}\bX) \,
\eEq
and, {\em locally}, $F_{\ome(\lam,\mathfrak{G}(\bar{\Xi})_{V})} = D_{H}M_{\ome(\lam,\mathfrak{G}(\bar{\Xi})_{V})}$,
with 
\beq
& M_{\ome(\lam,\mathfrak{G}(\bar{\Xi})_{V})}: \\ 
& J_{4s-1}\bY_{\zet} \ucar{\bX} VJ_{4s-1}\cA^{(r,k)} \to 
\Con_{2s}^{*}[\cA^{(r,k)}]\ten\Con_{2s-1}^{*}[\cA^{(r,k)}]\ten\Con_{0}^{*}[\cA^{(r,k)}]\wed 
(\owed{n-1}T^{*}\bX)\,.
\eeq
\eLm

\bDf
We call the global morphism $\bet(\lam,\mathfrak{G}(\bar{\Xi})_{V}) \byd E_{\ome(\lam,\mathfrak{G}(\bar{\Xi})_{V})}$ 
the {\em generalized} {\em Bianchi morphism} associated
with the Lagrangian $\lam$ and the variation vector field $\bar{\Xi}$.\END\eDf

\bRm
For any $(\bar{\Xi},\xi)\in \cA^{(r,k)}$, as a consequence of the gauge-natural invariance of the Lagrangian, 
the morphism $\bet(\lam,\mathfrak{G}(\bar{\Xi})_{V}) \equiv
\cE_{n}(\ome(\lam,\mathfrak{G}(\bar{\Xi})_{V}))$ is {\em locally} identically vanishing. 
We stress that these are just {\em local generalized Bianchi 
identities} \cite{Ber58}.\END
\eRm

Let $\mathfrak{K}$
be the {\em kernel} of
 $\cJ(\lam,\mathfrak{G}(\bar{\Xi})_{V})$. 
As a consequence of the considerations
above we have the following result \cite{PaWi03}.

\bTh
The generalized Bianchi morphism is globally vanishing for the variation vector field $\bar{\Xi}$ if and only 
if $\del^{2}_{\mathfrak{G}}\lam\equiv\cJ(\lam,\mathfrak{G}(\bar{\Xi})_{V})= 
0$, \ie if and only if
$\mathfrak{G}(\bar{\Xi})_{V}\in\mathfrak{K}$.
\eTh

From now on we shall write $\ome(\lam,
\mathfrak{K})$ to denote $\ome(\lam,\mathfrak{G}(\bar{\Xi})_{V})$ when $\mathfrak{G}(\bar{\Xi})_{V}$ belongs to $\mathfrak{K}$. Analogously
for $\bet$ and other morphisms. 

As we
already recalled, the  classical Jacobi equation for geodesics of a Riemannian manifold defines in 
fact the Riemann curvature tensor of the metric $g$; because of this it was suggested in \cite{Vari} that the 
second variation  and the generalized Jacobi equations define the 
`curvature' of any given variational principle. 
This general concept of
`curvature' takes, for example, a  particularly significant form in the case of generalized harmonic 
Lagrangians, giving rise to suitable `curvature tensors' which satisfy 
suitable `generalized Bianchi identities' \cite{Vari}. Following this guideline, we shall provide a definition of 
the curvature of a given gauge-natural invariant variational principle.

First of all let us make the following important consideration.

\bPr
For each $\bar{\Xi}\in \cA^{(r,k)}$ such that $\bar{\Xi}_{V}\in \mathfrak{K}$, we have
\bEq
\cL_{j_{s}\bar{\Xi}_{H}}\ome(\lam,
\mathfrak{K})=-D_{H}(-j_{s}\pounds_{\bar{\Xi}_{V}} 
\rfloor p_{D_{V}\ome(\lam,\mathfrak{K})}) \,.
\eEq
\ePr

\bPf
The horizontal splitting gives us $\cL_{j_{s}\bar{\Xi}}\ome(\lam,\mathfrak{K})=\cL_{j_{s}\bar{\Xi}_{H}}\ome(\lam,\mathfrak{K}) +
\cL_{j_{s}\bar{\Xi}_{V}}\ome(\lam,\mathfrak{K})$. Furthermore, $\ome(\lam,\mathfrak{K})\equiv - \pounds_{\bar{\Xi}} \rfloor
\cE_{n}(\lam) = \cL_{j_{s}\bar{\Xi}}\lam - d_{H}(-j_{s}\pounds_{\bar{\Xi}} 
\rfloor p_{d_{V}\lam}+ \xi \rfloor \lam)$; so that
\beq
\cL_{j_{s}\bar{\Xi}_{V}}\ome(\lam,\mathfrak{K})=\cL_{j_{s}\bar{\Xi}_{V}}\cL_{j_{s}\bar{\Xi}}\lam=\cL_{j_{s}[\bar{\Xi}_{V},
\bar{\Xi}_{H}]}\lam\,.
\eeq
 On the other hand we have 
\beq
\cL_{j_{s}\bar{\Xi}_{H}}\ome(\lam,\mathfrak{K}) = \cL_{j_{s}[\bar{\Xi}_{H},
\bar{\Xi}_{V}]}\lam=-\cL_{j_{s}\bar{\Xi}_{V}}\ome(\lam,\mathfrak{K})\,.
\eeq

Recall now that from the Theorem above we have $\bar{\Xi}_{V}\in
\mathfrak{K}$ if and only if $\bet(\lam,\mathfrak{K})=0$. 
Since 
\beq
& &\cL_{j_{s}\bar{\Xi}_{V}}\ome(\lam,\mathfrak{K})=
 - \pounds_{\bar{\Xi}_{V}} \rfloor \cE_{n}(\ome(\lam,\mathfrak{K})) 
+D_{H}(-j_{s}\pounds_{\bar{\Xi}_{V}} 
\rfloor p_{D_{V}\ome(\lam,\mathfrak{K})}) = \\
& &   = \bet(\lam,\mathfrak{K})+D_{H}(-j_{s}\pounds_{\bar{\Xi}_{V}} 
\rfloor p_{D_{V}\ome(\lam,\mathfrak{K})})\,,
\eeq
 we get the assertion.
\END
\ePf

It is easy to realize that, because of the gauge-natural invariance of the generalized
Lagrangian $\lam$, the new generalized Lagrangian $\ome(\lam,\mathfrak{K})$ is gauge-natural invariant too, \ie
$\cL_{j_{s}\bar{\Xi}}\,\ome(\lam,
\mathfrak{K})=0$.

Even more, we can state the following.
\bPr
Let $\bar{\Xi}_{V}\in
\mathfrak{K}$. We have
\bEq
\cL_{j_{s}\bar{\Xi}_{H}}\ome(\lam,
\mathfrak{K})=0\,.
\eEq
\ePr 

\bPf
In fact, when $\bar{\Xi}_{V}\in
\mathfrak{K}$, since $\cL_{j_{s}\bar{\Xi}_V}\ome(\lam,
\mathfrak{K})=[\bar{\Xi}_V,\bar{\Xi}_V]\rfloor \cE_n (\lam)+ \bar{\Xi}_V \rfloor\cL_{j_{s}\bar{\Xi}_V}\cE_n (\lam) =0$, we have 
\bEq
& &0=\cL_{j_{s}\bar{\Xi}}\ome(\lam,
\mathfrak{K})= \cL_{j_{s}\bar{\Xi}_V}\ome(\lam,
\mathfrak{K})+\cL_{j_{s}\bar{\Xi}_H}\ome(\lam,
\mathfrak{K})= \cL_{j_{s}\bar{\Xi}_H}\ome(\lam,
\mathfrak{K})\,.
\eEq
\QED
\ePf

As a quite relevant byproduct we get also the following (this result can be compared with \cite{CFT96}).

\bCr
Let $\bar{\Xi}_{V}\in
\mathfrak{K}$. We have the {\em covariant} conservation law
\bEq
D_{H}(-j_{s}\pounds_{\bar{\Xi}_{V}} 
\rfloor p_{D_{V}\ome(\lam,\mathfrak{K})})=0\,.
\eEq
\eCr

\bPf
\beq
& & 0=\cL_{j_{s}\bar{\Xi}_H}\ome(\lam,
\mathfrak{K})= -\bet(\lam,\mathfrak{K}) - \\
& &D_{H}(-j_{s}\pounds_{\bar{\Xi}_{V}} 
\rfloor p_{D_{V}\ome(\lam,\mathfrak{K})}) = \\ 
& & - D_{H}(-j_{s}\pounds_{\bar{\Xi}_{V}} 
\rfloor p_{D_{V}\ome(\lam,\mathfrak{K})})\,.
\eeq
\QED\ePf

\bDf
We define the covariantly conserved current 
\bEq
\cH(\lam,\mathfrak{K})=-j_{s}\pounds\,
\rfloor p_{D_{V}\ome(\lam,\mathfrak{K})}
\,,
\eEq
 to be a Hamiltonian form for
$\ome$ on the Legendre bundle $\Pi\equiv V^{*}(J_{2s}\bY_{\zet} \ucar{\bX}
VJ_{2s}\cA^{(r,k)})\wed (\owed{n-1}  T^{*}\bX)$  (in the sense of
\cite{MaSa00}).
\END
\eDf

Let $\Ome$ be the multisimplectic form on $\Pi$.
It is well known that every Hamiltonian form $\cH(\lam,\mathfrak{K})$ admits a canonical Hamiltonian connection
$\gam_{\cH}(\lam,\mathfrak{K})$ such that
$
\gam_{\cH}(\lam,\mathfrak{K})\rfloor\Ome = d\cH(\lam,\mathfrak{K})$. Let then  $\gam_{\cH}(\lam,\mathfrak{K})$ be the
corresponding Hamiltonian connection form (see
\cite{MaSa00}).

\bDf
We define the {\em curvature} of the given gauge-natural invariant variational principle to be the curvature of 
the Hamiltonian connection form $\gam_{\cH}(\lam,\mathfrak{K})$.\END
\eDf

We claim that the Euler--Lagrange and the generalized Jacobi
equations for $\lam$ can be related with the zero curvature
equations for the
Hamiltonian connection form $\gam_{\cH}(\lam,\mathfrak{K})$. According to
\cite{Vari,FrPa00,FrPa01}, they must be, in fact, the Lagrangian
counterpart of Hamilton
equations for the Lagrangian
$\ome$ obtained by contracting Euler--Lagrange equations with Jacobi
variation vector fields. Investigations are in progress and results will appear
elsewhere
\cite{FrPaWi05}.


\end{document}